\documentclass[a4paper,12pt]{article}
\usepackage{epsfig}
\usepackage{amssymb}
\usepackage{amsfonts}
\usepackage{amsmath}
\usepackage{euscript}
\usepackage{verbatim}
\usepackage{latexsym}
\usepackage{graphicx}

	\usepackage{graphicx}	
	\usepackage{subfigure}
	\usepackage{wrapfig}
	\usepackage[T1]{fontenc}
	\usepackage{mathtools}
	\usepackage{float}

\usepackage{tikz}
\usetikzlibrary{shapes.geometric, arrows,patterns,snakes}
\tikzstyle{ellip} = [ellipse, minimum width=3cm, minimum height=1cm,text centered, draw=black]

\newskip\humongous \humongous=0pt plus 1000pt minus 1000pt

\newif\ifdtup

\catcode`\@=11

\@addtoreset{equation}{section}

\def\@normalsize{\@setsize\normalsize{15pt}\xiipt\@xiipt
\abovedisplayskip 14pt plus3pt minus3pt%
\belowdisplayskip \abovedisplayskip
\abovedisplayshortskip \z@ plus3pt%
\belowdisplayshortskip 7pt plus3.5pt minus0pt}

\def\small{\@setsize\small{13.6pt}\xipt\@xipt
\abovedisplayskip 13pt plus3pt minus3pt%
\belowdisplayskip \abovedisplayskip
\abovedisplayshortskip \z@ plus3pt%
\belowdisplayshortskip 7pt plus3.5pt minus0pt
\def\@listi{\parsep 4.5pt plus 2pt minus 1pt
     \itemsep \parsep
     \topsep 9pt plus 3pt minus 3pt}}

\relax

\catcode`@=12

\catcode`\@=11

\def\section{\@startsection{section}{1}{\z@}{3.5ex plus 1ex minus
   .2ex}{2.3ex plus .2ex}{\large\bf}}

\def\SymBoxes#1#2#3#4{\newdimen\un@t \un@t#3%
\raisebox{#1}{\rule{#2\un@t}{#4}\hskip-#2\un@t% lower horizontal
\@tempdimb\un@t \advance\@tempdimb by-#4\@tempcntb#2\relax%
\@whilenum{\@tempcntb>0}\do{%                         % #2 vertical lines
\rule{#4}{\un@t}\hskip\@tempdimb \advance\@tempcntb by\m@ne}%
\hskip-#2\un@t \rule[\un@t]{#2\un@t}{#4}%
\rule[\un@t]{#4}{#4}\hskip-#4%             % upper horizontal line
\rule{#4}{\un@t}}\hskip-#4}                % rightest vertical line
\begin{document}
%\begin{letter}{~}

%%%%%%Define some new commands and  macros
\newcommand{\beq}{\begin{equation}}
\newcommand{\eeq}{\end{equation}}
\newcommand{\bea}{\begin{eqnarray}}
\newcommand{\eea}{\end{eqnarray}}
\newcommand{\beas}{\begin{eqnarray*}}
\newcommand{\eeas}{\end{eqnarray*}}
\newcommand{\defi}{\stackrel{\rm def}{=}}
\newcommand{\non}{\nonumber}
\newcommand{\bquo}{\begin{quote}}
\newcommand{\enqu}{\end{quote}}
%%%%%%%%%%%%%%%%
\renewcommand{\(}{\begin{equation}}
\renewcommand{\)}{\end{equation}}
%%%%%%%%%%%%%%%%%%%%%%%%%%%%%%%%%% definitions
\def \eqn#1#2{\begin{equation}#2\label{#1}\end{equation}}

\def\bA{\bar{A}}
\def\ba{\bar{a}}
\def\e{\epsilon}
\def\calM{{\mathcal M}}
\def\cL{\mathcal{L}}
\def\cW{\mathcal{W}}
\newcommand{\calL}[1] {\mathcal{L}^{(#1)}}
\newcommand{\calW}[1] {\mathcal{W}^{(#1)}}
\newcommand{\mui}[1] {\mu^{(#1)}}
\newcommand{\nui}[1] {\nu^{(#1)}}
\newcommand{\bcalL} {\overline{\cL}}
\newcommand{\bcalW} {\overline{\cW}}
\def\bmu{\overline{\mu}}
\def\bnu{\overline{\nu}}
\newcommand{\ei}[1] {\epsilon^{(#1)}}
\newcommand{\si}[1] {\sigma^{(#1)}}
\def\pp{\partial_{\phi}}
\def\pt{\partial_t}
\def\rme{\mathrm{e}}
\def\Tr{ \hbox{\rm Tr}}
\def\H{ \hbox{\rm H}}
\def\HE{ \hbox{$\rm H^{even}$}}
\def\HO{ \hbox{$\rm H^{odd}$}}
\def\K{ \hbox{\rm K}}
\def\Im{ \hbox{\rm Im}}
\def\Ker{ \hbox{\rm Ker}}
\def\const{\hbox {\rm const.}}
\def\o{\over}
\def\im{\hbox{\rm Im}}
\def\re{\hbox{\rm Re}}
\def\bra{\langle}
\def\ket{\rangle}
\def\Arg{\hbox {\rm Arg}}
\def\Re{\hbox {\rm Re}}
\def\Im{\hbox {\rm Im}}
\def\exo{\hbox {\rm exp}}
\def\diag{\hbox{\rm diag}}
\def\longvert{{\rule[-2mm]{0.1mm}{7mm}}\,}
\def\a{\alpha}
\def\dag{{}^{\dagger}}
\def\tq{{\widetilde q}}
\def\p{{}^{\prime}}
\def\W{W}
\def\N{{\cal N}}
\def\calB{\mathcal{B}}
\def\hsp{,\hspace{.7cm}}

\def\br{\nonumber\\}
\def\bcalL{\bar{\mathcal{L}}}
\def\bcalW{\bar{\mathcal{W}}}
\def \eqn#1#2{\begin{equation}#2\label{#1}\end{equation}}

\newcommand{\M}{\ensuremath{\mathcal{M}}
                    }
\newcommand{\oc}{\ensuremath{\overline{c}}}
\begin{titlepage}
\begin{flushright}
CHEP XXXXX
%ULB-TH/09-10\\
%hep-th/yymmnnn\\
\end{flushright}
\bigskip
\def\thefootnote{\fnsymbol{footnote}}

\begin{center}
{\Large
{\bf A Neumann Boundary Term for Gravity \\ 
\vspace{0.1in} 
}
}
\end{center}

\bigskip
\begin{center}
{%\large Chethan KRISHNAN$^a$\footnote{\texttt{chethan.krishnan@gmail.com}} and  
Chethan KRISHNAN$^a$\footnote{\texttt{chethan.krishnan@gmail.com}}, Avinash RAJU$^a$\footnote{\texttt{avinashraju777@gmail.com}}}
\vspace{0.1in}

\end{center}

\renewcommand{\thefootnote}{\arabic{footnote}}

\begin{center}
%\vspace{0.2cm}

$^a$ {Center for High Energy Physics,\\
Indian Institute of Science, Bangalore 560012, India}\\

\end{center}

\noindent
\begin{center} {\bf Abstract} \end{center}
The Gibbons-Hawking-York (GHY) boundary term makes the Dirichlet problem for gravity well defined, but no such general term seems to be known for Neumann boundary conditions. In this paper, we view Neumann {\em not} as fixing the normal derivative of the metric (``velocity") at the boundary, but as fixing the functional derivative of the action with respect to the boundary metric (``momentum"). This leads directly to a new boundary term for gravity: the trace of the extrinsic curvature with a specific dimension-dependent coefficient. In three dimensions this boundary term reduces to a ``one-half" GHY term noted in the literature previously, and we observe %in flat space and AdS$_3$, 
that our action %in the metric language 
translates precisely to the Chern-Simons action with no extra boundary terms. % in the first order formulation. 
In four dimensions the boundary term vanishes, giving a natural Neumann interpretation to the standard Einstein-Hilbert action without boundary terms. We argue that in light of AdS/CFT, ours is a natural approach for defining a ``microcanonical" path integral for gravity in the spirit of the (pre-AdS/CFT) work of Brown and York.

%As a simple extension, we also write down the general Robin boundary term for gravity.

%In the 't Hooft limit, where $\lambda$ is fixed and $N \rightarrow \infty$,  gauge theory effectively becomes classical. Classical theories can thermalize, so we consider seriously the possibility that in the dual gravity picture, one should be able to understand black hole formation by studying tree level string theory. We speculate that global event horizons do not form under gravitational collapse when gravity has non-localities  

\vspace{1.6 cm}
\vfill

\end{titlepage}

\setcounter{footnote}{0}

%%%%%%%%%%%%%%%%%%%%%%%%%%%%%%%%%%%%%%%%%%%%%%%%%%%%%%%%%%%%%%%%%%%%%%%%%%%%%%%%%%%%%%%%%%%%%%
%%%%%%%%%%%%%%%%%%%%%%%%%%%%%%%%%%%%%%%%%%%%%%%%%%%%%%%%%%%%%%%%%%%%%%%%%%%%%%%%%%%%%%%%%%%%%%

\section{Introduction}
%All comments about Neumann (including Headrick, McNees, Compere-Marolf, ...)

To derive an equation of motion from an action using a variational principle (see e.g., \cite{Hint}), we need to make sure that the boundary terms arising from the variation vanish. For two-derivative theories, the boundary terms that arise from the variation of the action typically contain variations of both the field and its normal derivative at the boundary. Holding both fixed at the boundary will trivially get rid of these terms, but it will also remove most of the interesting dynamics. Instead, what one tries to do is to add {\em boundary terms} to the action such that the total boundary variation after the addition of these new terms depends either only on the field variation, or only on the normal derivative variation. When we can find a boundary term to accomplish this, we say that we have a Dirichlet problem (in the former case) or a Neumann problem (in the latter). In such a situation, we have a well-defined variational problem upon demanding that the field (for Dirichlet) or its normal derivative (for Neumann) be held fixed at the boundary.

In the case of gravity, it has been known since \cite{GH, Y} that there exists a boundary term one can add to the bulk Einstein-Hilbert action (\ref{EH}) to make the Dirichlet problem for the metric well-defined. This is the Gibbons-Hawking-York (GHY) term (\ref{GHY}). But the Neumann problem as we have stated in the above paragraph does not seem to be well-defined for gravity. This is because, (to the best of our knowledge) no one has written down a boundary term to be added to the Einstein-Hilbert action such that {\em just} holding the normal derivatives fixed at the boundary kills all boundary terms that arise from the variation of the total action \footnote{See for example \cite{Stefan} for some explicit expressions for the variations in the context of three dimensions; the structure in general dimensions is similar.}. This is striking, since the GHY term has been around for forty years. 

In this paper, we will take an alternative view on the Neumann problem. Instead of holding the normal derivative of the metric fixed at the boundary, we will seek a variational problem where the functional derivative of the action with respect to the boundary metric is held fixed. In the case of classical particle mechanics, these two formulations are equivalent because the former corresponds to holding the velocity fixed at the boundary, whereas the latter corresponds to holding the momentum fixed. But we will see that in the case of gravity, the latter formulation results in a drastic simplification, and we can indeed write down a boundary term that makes this type of a Neumann problem well-defined.

The basic idea here is very natural, and indeed unavoidable in a diffeomorphism invariant theory: we need a covariant notion of the Neumann boundary {\em condition} before we can talk about a covariant Neumann boundary {\em term}. Holding the normal derivative fixed is {\em not} a covariant notion in a generic spacetime, whereas our approach is. Our approach can also be understood in terms of a Legendre transform as we will explore in various contexts in \cite{Pavan, Bala, Robin, Chethan}.

\section{Dirichlet problem}

We will start by reviewing the Dirichlet problem for gravity. Everything in this section is well-known, but we want to write the Dirichlet action in a form that is suitable for moving to Neumann. %The Einstein-Hilbert action in D-dimensions with a cosmological term is given by

For a well defined variational problem with Dirichlet boundary conditions, we need to add a boundary piece to the Einstein-Hilbert action so that when the boundary metric $h_{ij}$ is held fixed and when the bulk equations of motion hold, the variation of the total action becomes zero. 
%whose variation is given by  \begin{equation} \delta S_{GHY} = \frac{1}{\kappa}\int_{\partial {\cal M}}d^{D-1}y \sqrt{|h|}\varepsilon \left(\delta K + \frac{1}{2}K h^{ij}\delta h_{ij} \right). \end{equation}
Demanding this results in the total gravitational action for the Dirichlet problem  
\begin{eqnarray}
S_{D} &=& S_{EH} +  S_{GHY}
\end{eqnarray}
where 
\begin{equation}
S_{EH} = \frac{1}{2\kappa}\int_{{\cal M}}d^{D}x \sqrt{-g}(R-2\Lambda)\label{EH}
\end{equation}
and the boundary piece is the GHY term
\begin{equation}
S_{GHY} = \frac{1}{\kappa}\int_{\partial {\cal M}}d^{D-1}y \sqrt{|h|} \varepsilon K. \label{GHY}
\end{equation}
Here $\kappa=8\pi G$ %Variation yields \begin{eqnarray} \delta S_{EH} = \frac{1}{2\kappa}\int_{{\cal M}}d^{D}x \sqrt{-g}(G_{ab}+\Lambda g_{ab})\delta g^{ab}+ \nonumber \hspace{0.4in}\\  \hspace{0.2in}-\frac{1}{\kappa}\int_{\partial {\cal M}}d^{D-1}y \sqrt{|h|} \varepsilon \left( \delta K + \frac{1}{2}K^{ij}\delta h_{ij} \right) \end{eqnarray}
and $G_{ab} = R_{ab}-\frac{1}{2}R g_{ab}$. $h_{ij} \equiv g_{ab}e^{a}_{i}e^{b}_{j}$ refers to the induced metric on the boundary $\partial {\cal M}$ and $e^{a}_{i} = \frac{\partial x^a}{\partial y^i}$ are the projectors relating the bulk coordiantes $x^a$ to boundary coordinates $y^i$. $\varepsilon = \pm 1$ for time/space-like boundary. %and $\varepsilon =1$ for time-like boundary. 
Null boundaries are not considered here. If $n^{a}$ is the unit normal vector field on the boundary hypersurface, the extrinsic curvature is 
%\begin{equation}
$K_{ij} = \frac{1}{2}(\nabla_a n_b + \nabla_b n_a) e^{a}_{i}e^{b}_{j}$,
%\end{equation}
and $K = h^{ij}K_{ij}$.

The variation of $S_D$ is
\begin{eqnarray}
\delta S_{D}=\frac{1}{2\kappa}\int_{{\cal M}}d^{D}x \sqrt{-g}(G_{ab}+\Lambda g_{ab})\delta g^{ab}+ \label{Dvar} \\ \hspace{0.4in} -\frac{1}{2\kappa}\int_{\partial {\cal M}}d^{D-1}y \sqrt{|h|} \varepsilon  \left( K^{ij} - K h^{ij}  \right)\delta h_{ij}\nonumber 
\end{eqnarray}
This means that the action $S_{D}$ is stationary under arbitrary variations of the metric in the bulk provided we hold $h_{ij}$ fixed at the boundary (assuming the bulk equations of motion hold). Just the Einstein-Hilbert piece alone, without the GHY term, does {\em not} have this property.  This is what identifies $S_D$ as the Dirichlet action. 

\section{Neumann problem}

Before moving on to the discussion of Neumann, we define:
\begin{equation}
\pi^{ij} \equiv \frac{\delta S_{D}}{\delta h_{ij}} = -\frac{\sqrt{|h|}}{2\kappa} \varepsilon (K^{ij}-K h^{ij})
\end{equation}
The point of this is that the variation of the Dirichlet action (\ref{Dvar}) can now be written in the suggestive form
\begin{eqnarray}\label{variation}
\delta S_{D} = \hspace{2.5in} \\  \frac{1}{2\kappa}\int_{\cal M}d^{D}x \sqrt{-g}(G_{ab}+\Lambda g_{ab})\delta g^{ab}  +\int_{\partial {\cal M}}d^{D-1}y\  \pi^{ij} \delta h_{ij}\nonumber
\end{eqnarray}

We want to write down a variational principle where instead of holding the metric $h_{ij}$ fixed at the boundary, we can hold $\pi^{ij}$ fixed \footnote{Note that in particle mechanics, holding $\dot q$ fixed and holding $p$ fixed at the boundary are identical because of the simple ${\dot q}^2$ form of the kinetic term, which guarantees that $p(q,\dot q)=\dot q$. Note that both in particle mechanics as well as in our case for gravity, we are using the suggestive symbols $p$ and $\pi^{ij}$, but thinking of them as functions of $(q, \dot q)$ and $(h_{ij}, \partial_a h_{ij})$ respectively.}. 
It is easy to see from (\ref{variation}) that this can be easily accomplished by adding yet another term to the Dirichlet action of the previous section. The form suggested by (\ref{variation}) is
\begin{equation}
S_{N} = S_{D} -\int_{\partial {\cal M}} d^{D-1}y \ \pi^{ij}h_{ij}
\end{equation}
It is trivial now to check that the variation of $S_{N}$ is
\begin{eqnarray}
\delta S_{N} = \hspace{2.5in} \\ \frac{1}{2\kappa}\int_{{\cal M}}d^{D}x \sqrt{-g}(G_{ab}+\Lambda g_{ab})\delta g^{ab} - \int_{\partial {\cal M}}d^{D-1}y\  \delta \pi^{ij} h_{ij}\nonumber
\end{eqnarray}
guaranteeing that holding $\pi^{ij}$ fixed at the boundary yields a well-defined variational problem. This is the main observation of this paper. 

Explicitly, our Neumann action is given by
\begin{eqnarray}
S_{N} & = & S_{EH}+S_{N_b} \\
& \equiv & S_{EH} %\frac{1}{2\kappa}\int_{{\cal M}}d^{D}x \sqrt{-g}(R-2\Lambda) 
+ \frac{(4-D)}{2\kappa}\int_{\partial {\cal M}}d^{D-1}y \sqrt{|h|}\varepsilon K \label{NeumannA}
\end{eqnarray}
In the rest of this paper, we will discuss features, applications and extensions of this action. We will see that it is natural one to consider from many different angles.
%We believe this is a new result, and since it seems rather surprising that this simple result has not been noted the literature since the 40 years since the GHY papers, we will comment about some related work in the literature and how their work is different from ours. 

\section{D=2, 3, 4}

In two dimensions, our boundary term is identical to the GHY boundary term. This might seem puzzling, but is easy to understand by writing the metric in conformal gauge. The only degree of freedom is the conformal factor in the metric $ ds^2=e^{2 \phi (t,r)} (-dt^2+dr^2)$
and in terms of $\phi$, the Einstein-Hilbert action takes the form 
\begin{eqnarray}
S_{EH} \sim \int dt\ dr (\ddot \phi-\phi'')
\end{eqnarray}
Variation of this results in boundary terms of the form $\int dt \ \delta \phi'|_{\Lambda}$. The GHY term is designed precisely to cancel this when varied, and unlike in higher dimensions this form is trivial enough that its variation kills off the entire boundary piece. We have taken the boundary to be timelike and put it at $r=\Lambda$ for definiteness, but the discussion is obviously analogous for a spacelike boundary. 

In more than two dimensions, the Dirichlet and Neumann boundary terms are always distinct.

In three dimensions, our Neumann boundary term reduces to $S_{N_b}=\frac{1}{2\kappa}\int_{\partial {\cal M}}d^{2}y \sqrt{|h|}\varepsilon K$, which is ``one-half" the Gibbons-Hawking term.  Remarkably, this boundary term has previously appeared in the literature for other reasons both in the flat case \cite{Stefan} as well as  in AdS \cite{Banados, Olea}. %The standard Gibbons-Hawking term in 2+1 dimensions translates to a boundary term of the form  (see for example, \cite{Apolo}) 

Especially in light of holography in AdS$_3$ and flat 2+1 dimensional space, there is much to be said here. But we will leave that for future work, and make only one brief comment: the Einstein-Hilbert action leads to a boundary piece
\begin{eqnarray}
S^{CS}_{B}\sim \int_{\partial {\cal M}} {\rm Tr} (A \wedge \bar A)
\end{eqnarray}
in the Chern-Simons formulation (see eg., \cite{Kiran, KRNEW} for an elementary discussion of Chern-Simons gravity in 2+1 dimensions; we work with the AdS case for concreteness). It is straightforward to check that the ``one-half" Gibbons-Hawking term precisely gets rid of this boundary piece \footnote{See for example \cite{Apolo}, the discussion at the beginning of section 2.1. The ``one-half" GHY term is not discussed there, but our claims about it follow trivially from the expressions there.}. Therefore in 2+1 dimensions, our boundary term makes the metric formulation of gravity translate precisely into the bulk Chern-Simons action, with {\em no} boundary term at all. A natural choice of boundary conditions \cite{Banados} in the Euclidean geometry is then to set $A_z=0$ or $A_{\bar z}=0$ at the boundary (here $z \sim (t+ix)$ where $t$ and $x$ are boundary coordinates). These boundary conditions are satisfied by AdS$_3$ and the BTZ black hole.  

In four dimensions, our boundary term identically vanishes. Thus, we come to a perhaps surprising conclusion: standard Einstein-Hilbert gravity in four dimensions, {\em without} boundary terms, has an interpretation as a Neumann problem\footnote{The variation of the $D=4$ Einstein-Hilbert action in this form, without emphasis of the Neumann interpretation, can be found (say) in eqn. (18) of \cite{york-old}. We suspect that this ``accidental'' triviality in $D=4$ is one reason why the Neumann term was missed in generic $D$.}. 

\section{Microcanonical Gravity}

In \cite{BY}, a microcanonical definition of the gravitational path integral was proposed. The basic idea there was to add new boundary terms to the gravitational field, so that the energy surface \footnote{The term ``surface" arises because \cite{BY} work in four dimensions, but the approach is more general.} density and the momentum surface density are held fixed at the boundary, in the definition of the variational principle. One can then use this action to define a microcanonical path integral for gravity which has some pleasing properties. 

Even though \cite{BY} does not emphasize it, it is easy to see that the surface energy/momentum densities that they hold fixed are just some of the components of the energy momentum surface density, which is the quantity we have held fixed in defining our Neumann variational problem. The approach of \cite{BY} results in the somewhat awkward action, eqn. (3.13) in their paper. However, despite this, since charges are best defined globally in general relativity, we feel that the approach of \cite{BY} is a very interesting one. This is one of the motivations for defining a more ``covariant" variational problem where it is the whole stress-energy tensor density on the boundary  that is held fixed. Happily, this also turns out to have a Neumann interpretation. 

The work of \cite{BY} was before the era of AdS/CFT, and in hindsight, we believe this approach is an even more interesting one in the AdS/CFT context. Fixing the boundary stress energy tensor gives us a natural definition for a microcanonical approach to AdS/CFT. We will be reporting on this in greater detail elsewhere \cite{Bala}, but we briefly outline some of the results here. 

The natural boundary stress tensor in AdS/CFT is the one introduced by \cite{BK} (see also, for example \cite{Skenderis}). These stress tensors are obtained by adding further boundary terms (which have a natural interpretation as counter-terms in the holographic language) while demanding that the on-shell action be finite \footnote{The original action has a bulk IR divergence.}. This gives rise to the finite {\em renormalized boundary stress tensors} found in \cite{BK, Skenderis}. 

Together with the results of the current paper, this suggests that a natural object to hold fixed while doing a Neumann variation in the AdS/CFT context is the renormalized stress tensor density, not the bare stress tensor density. Indeed, it is possible to show \cite{Bala} that one can add appropriate counter-terms to our Neumann action (\ref{NeumannA}) so that:

(a) the variation of the total action leads to the {\em renormalized} stress tensor of \cite{BK, Skenderis} (up to an ambiguity in odd dimensions \cite{Bala}), 

(b) the variational principle is well-defined when one holds this renormalized boundary stress tensor density fixed, and 

(c) the total on-shell action is finite. \\
\noindent
A detailed discussion of these issues in various dimensions will be presented in a companion paper \cite{Bala} in the standard language of holographic renormalization. There, we will also argue that AdS/CFT is the natural context for discussing, extending and finding applications for the (stress-tensor version of the) microcanonical path integral approach of \cite{BY} \footnote{For example, note that in the work of \cite{BY} there was no guarantee that the action was finite once you took the boundary to infinity.}. 

In this letter, we will merely settle for presenting the actions satisfying the above conditions in AdS$_3$ and AdS$_4$. Interestingly, in AdS$_3$ the action is already finite (upto a logarithmic counter-term) with our choice of boundary term, and it leads \footnote{Up to an ambiguity that affects odd-dimensional AdS spaces that we will elaborate in \cite{Bala}. There we will also discuss the possibility of the logarithmic counter term that is related to the trace anomaly \cite{Skend0}. The standard black hole solutions do not require it in order to have a finite action, so we will not emphasize it here.} to the renormalized stress tensor of \cite{BK, Skenderis}: we do not need any further counter terms. Indeed, some of these facts are responsible for the ``one-half" GHY term having been noted previously in the literature.

In four dimensions however, we discover a new counter-term:
\begin{equation}
S^{ct}_N = \frac{1}{\kappa}\int d^3 y\; \sqrt{|h|}\left(1- \frac{1}{4}R[h] \right)
\end{equation}
(The AdS radius $\ell$ has been set to one.)
Note that the specific coefficients differ from what one finds in the counter-terms in \cite{BK}, but the stress tensor resulting from the total action is finite and identical to that in \cite{BK}. This is because we are adding this counter-term to (\ref{NeumannA}), not to the standard Dirichlet action. Satisfyingly, the on-shell action also turns out to be finite in this approach.  %A detailed discussion of these issues in various dimensions 

%\section{Robin Boundary Conditions} Robin boundary condition refers to fixing a linear combination of the field and its derivative fixed at the boundary. The analogous thing to do in our approach would be to hold \begin{eqnarray} h_{ij}+ \alpha \ \pi_{ij}  \end{eqnarray} fixed, where $\alpha$ is some real constant. 

\section{Discussion}

A standard Neumann boundary term for gravity, where one can obtain a well-defined variational problem by simply holding the normal derivative of the metric fixed at the boundary, is not known for Einstein-Hilbert gravity to the best of our knowledge. Occasionally statements of the form $K_{ij}=0$  or $K_{ij}-\alpha h_{ij}=0$ are considered as Neumann or Robin (mixed) boundary conditions in the literature. But note that these are in fact far stronger conditions than the usual Neumann or Robin boundary conditions, which merely say that these quantities are fixed, {\em not} that they are zero. Indeed, these conditions put {\em differential} constraints on the boundary surface (surfaces which admit conditions of this type are sometimes called totally umbilical surfaces \cite{Zanelli}). In hindsight, our approach provides a natural explanation why some of these boundary conditions had nice properties: they arise as natural further restrictions on the Neumann boundary condition, which puts restrictions on the boundary stress tensor (or equivalently, the extrinsic curvature). In this context, we suspect that it should be possible to define a Neumann boundary condition for higher derivative gravity theories \cite{Hint} as well, by setting the functional derivative of the action with respect to the boundary metric fixed.

In the context of AdS/CFT some papers with a loosely Neumann flavor have appeared, one of the more visible ones being \cite{Compere}. What they do is to treat the bulk path integral (or action, when one is working semi-classically) as a functional of the boundary metric and then integrate over the boundary metric to define a new path integral. This has its interest, but its connection to the standard Neumann problem is not immediate. See \cite{others} for other various related discussions on boundary aspects of general relativity.

We believe that what we have considered in this paper is a natural version of the Neumann condition for gravity: our approach reduces to the standard Neumann problem when applied to particle mechanics, and it does not put any differential constraints the boundary surface. It also puts various curiosities in the literature on boundary terms, in context. Finally, it also leads to a natural definition of the microcanonical path integral for AdS gravity as will be elaborated elsewhere \cite{Bala}.

%We have searched (to the limits of our googling ability) to see what has been said about this in the literature, and we will summarize what we have found to contrast our approach in this paper.

{\bf Comment added:} After this preprint appeared on the arXiv, more work has been done developing this set of ideas further. These include \cite{Pavan} which studies a semi-classical path integral with Neumann boundary conditions and uses it to reproduce the thermodynamics of horizons, \cite{Bala} that applies this boundary term in the context of the AdS/CFT correspondence, and \cite{Robin} which does a slight generalization of the Neumann boundary term to allow Robin boundary conditions for general relativity. A paper \cite{Chethan} that discusses the conditions under which Neumann boundary conditions lead to normalizable fluctuations in AdS and explores its connection to conformal gravity on the boundary, is about to appear.

\section*{Acknowledgement}
CK thanks Geoffrey Compere for correspondence, and Daniel Grumiller (and through him Friedrich Scholler) for checking a preliminary draft of this manuscript, and for encouraging comments. We thank P. N. Bala Subramanian for collaboration on \cite{Bala}.

\end{document}